\documentclass[aps,llncs,preprint]{revtex4-2}
\usepackage{graphicx}
\usepackage{amsmath}
\usepackage{float}
\begin{document}

\title{Anisotropic superconductivity in ZrB$_{12}$ near critical Bogomolnyi point} 

\author{Soumya Datta$^1$, Sandeep Howlader$^1$, Arushi$^2$, Ravi Prakash Singh$^2$, and Goutam Sheet$^1$}

\email{goutam@iisermohali.ac.in}

\affiliation{$^1$Department of Physical Sciences, Indian Institute of Science Education and Research Mohali, Sector 81, S. A. S. Nagar, Manauli, PO 140306, India}

\affiliation{$^2$Department of Physics, Indian Institute of Science Education and Research Bhopal, Bhopal 462066, India}

\begin{abstract}
	
The superconductors with the Ginzburg-Landau (G-L) parameter ($\kappa$) $\sim1/\sqrt{2}$ exist near a critical Bogolomonyi (B) point where they show inter-type domains between type-I and type-II superconductivity. While such physics is well understood for isotropic superconductors, the experimental investigation of the physics of anisotropic superconductors near a critical B-point remains an unattained goal mainly due to the unavailability of model material systems. Theoretically, such superconductors are expected to show type-I or type-II behaviour for definite directions of an applied magnetic field. Here, from directional point-contact Andreev reflection spectroscopy and field-angle dependent ac magnetic susceptibility measurements, we show that ZrB$_{12}$ is an anisotropic superconductor, and it exhibits field-direction dependent type-I and type-II behavior. These observations match remarkably well with the theoretical expectations for an anisotropic superconductor near a critical B-point. Therefore, our results project ZrB$_{12}$ as a model material system where the physics of inter-type anisotropic superconductivity can be explored experimentally.

\end{abstract}

\maketitle

\section{Introduction}


In the phenomenological Ginzburg-Landau (G-L) theory\cite{landau1950theory,ginzburg2009theory}, the conventional superconductors are categorized into two distinct types according to their magnetic properties. The type-I superconductors are ideally diamagnetic, while type-II superconductors allow partial penetration of the external magnetic field in the form of single-quantum vortices. In the latter case, if the density of the vortices is sufficiently large, they form an Abrikosov lattice\cite{abrikosov1957magnetic,abrikosov2004nobel}. These two types are regularly described and distinguished by G-L parameter $\kappa$, which is the ratio between magnetic penetration depth $\lambda$ and superconducting coherence length $\xi$ of a superconductor\cite{tinkham2004introduction}. A superconductor behaves like a type-I if $\kappa < 1/\sqrt{2}$, and a type-II if $\kappa > 1/\sqrt{2}$. This classification, however, does not work if for a superconductor $\kappa \sim 1/\sqrt{2}$\cite{krageloh1969flux,essmann1971observation,aston1971intermediate,jacobs1971interaction,auer1973magnetic,luk2001theory}. On the other hand, the G-L formalism is strictly valid near the superconducting critical temperature ($T \sim T_c$). Hence, this classification between type-I and type-II superconductors can be achieved only near $T_c$. In the hypothetical phase space ($\kappa$, $T$), this critical point (1/$\sqrt{2}$, $T_c$) is referred to as Bogomolnyi point (B-point)
\cite{bogomol1976stability,jacobs1979interaction}. A fundamental difference between type-I and type-II superconductors is that the vortex-vortex interaction is attractive in the former but repulsive in the latter\cite{jacobs1979interaction}. However, precisely at the B-point, the interaction between the vortices completely disappears, and an infinite degeneracy of arbitrary flux configurations exists\cite{bogomol1976stability,jacobs1979interaction}. Deeper in the superconducting state, when $T < T_c$, the B-point spreads over a finite interval of $\kappa$ values\cite{vagov2016superconductivity,cordoba2016between,wolf2017bcs,wolf2017vortex}, and the standard description of type-I and type-II superconductivity fails\cite{jacobs1971interaction,auer1973magnetic,luk2001theory}. The superconductors falling in this category are referred to as `inter-type (IT) superconductors'. Here, the vortex-vortex interaction becomes non-monotonic. Consequently, attractive long-range interactions and repulsive short-range interactions both become simultaneously possible. Since the range of $\kappa$ over which the B-point spreads is temperature-dependent, a temperature-dependent transition from the usual type-I to a type-II superconducting phase is expected in such systems. Apart from that, the B-point being infinitely degenerate, the superconductor also becomes sensitive to other internal and external parameters like system geometry, impurities, applied current, and external magnetic field. Due to this sensitivity, the magnetic properties of an IT superconductor can be externally manipulated to exhibit some exotic behaviours\cite{weber1978transition,sporna1979anisotropy,koike1980superconductivity,moser1982superconductive,sauerzopf1987anisotropy,weber1989magnetization,de2017threshold,zhang2021full}. Based on some recent experimental findings, ZrB$_{12}$ is believed to be one example of such IT superconductors\cite{wang2005specific,ge2014direct,ge2017paramagnetic,biswas2020coexistence}. However, the above description of inter-type superconductivity is valid for isotropic superconductors. In this work, we report anisotropic superconductivity in ZrB$_{12}$, where the description of inter-type domain warrants re-investigation.\\


ZrB$_{12}$ has an fcc structure of space group \textit{Fm3m}\cite{post1952crystal,kennard1983zirconium,leithe2002refinement}, and it superconducts at $T_c$ $\sim$ 5.85 K\cite{matthias1968superconductivity,gasparov2004electron,daghero2004andreev,lortz2005specific,wang2005specific,khasanov2005anomalous,gasparov2005electron,gasparov2006two}. The optical phonon modes, which are associated with the internal motion of the Zr atoms inside the boron cage, are responsible for the superconductivity in this material\cite{fisk1971superconducting}. The band structure calculation revealed that the Fermi level of ZrB$_{12}$ is located at an extended flat plateau in its electronic density of states (DOS), which makes the superconducting phase highly stable under any perturbations like chemical impurities or crystal defects\cite{shein2003band}. As we will see later, this quality of ZrB$_{12}$ was particularly beneficial for our point contact Andreev reflection study. From the same calculation\cite{shein2003band}, it was also revealed that the Fermi surface of ZrB$_{12}$ consists of an open sheet with hole characters and a quasi-spherical sheet with electron characters. These structures were further verified from optical\cite{teyssier2007optical} and de Haas-van Alphen studies\cite{gasparov2008study}.\\


Despite numerous theoretical and experimental studies reported on ZrB$_{12}$, significant disagreements still persist in the literature regarding the superconducting properties of this system. Compared to an ideal BCS superconductor, the unusual behaviours of ZrB$_{12}$ can be broadly categorized in to three different types. In our present report, we will try to shine light on each one of these. The most widely studied feature of superconducting ZrB$_{12}$ is its electron-phonon coupling. Based on several independent bulk sensitive experimental studies like temperature-dependent resistivity, critical field, specific-heat, thermal-expansion experiments, etc., ZrB$_{12}$ was thought to be in the weak coupling regime\cite{daghero2004andreev,tsindlekht2004tunneling,lortz2005specific}. On the other hand, a number of experiments, like point contact spectroscopy, tunneling spectroscopy, de Haas-van Alphen effect, etc. indicated otherwise and pointed towards a strong coupling superconductivity in this material\cite{daghero2004andreev,tsindlekht2004tunneling,gasparov2004electron,teyssier2007optical,gasparov2008study,girovsky2010strong}. In order to explain the differences in the bulk and surface superconducting properties, it was proposed that this material has enhanced surface characteristics\cite{tsindlekht2004tunneling,khasanov2005anomalous}. The second issue is regarding the description of the superconducting order parameter. From the temperature dependence of the penetration depth study, an indication of \textit{d}-wave pairing symmetry was reported in ZrB$_{12}$\cite{gasparov2005electron}. The possibility of two-gap superconductivity was also argued in this material\cite{gasparov2006two,sluchanko201110,bolotina2021checkerboard}. However, detailed specific-heat, resistivity, magnetic susceptibility and thermal-expansion experiments provided evidence against such claims and supported the idea of single gap BCS superconductivity in ZrB$_{12}$\cite{lortz2005specific,wang2005specific}. Moreover, all the energy-resolved spectroscopic measurements performed on ZrB$_{12}$ so far, supported single gap s-wave symmetry only\cite{daghero2004andreev,tsindlekht2004tunneling}. The third unusual feature of superconducting ZrB$_{12}$, which is also the most relevant one in our present analysis, is its magnetic behaviour. Based on various bulk sensitive measurements, Lortz \textit{et al.}\cite{lortz2005specific} and Wang \textit{et al.}\cite{wang2005specific} reported that ZrB$_{12}$ undergoes a transition from type-I superconductivity near $T_{c}$ to a type-II behaviour below $\sim$ 4.6 K. That provided the first evidence that ZrB$_{12}$ could be classified in the special class of type-II/I superconductor. The reported G-L parameter ($\kappa$) $\sim$ 0.65\cite{wang2005specific} being close to the border value 1/$\sqrt{2}$ also supported the argument. Scanning Hall probe microscopy study\cite{ge2014direct} reported a vortex pattern transition in a ZrB$_{12}$ across the type-II and type-II/I phases. The paramagnetic Meissner effect study\cite{ge2017paramagnetic} further reported that vortex clusters mediate the expulsion and penetration of flux at low temperatures. Recently, muon spin rotation measurements\cite{biswas2020coexistence} performed on ZrB$_{12}$ revealed that the type-I and type-II behaviours coexist within a finite temperature range simultaneously. This latest observation contrasts with the previous idea of a crossover from type-I to type-II superconductivity at a fixed temperature. Here, the authors argued about the controversial type-1.5 regime\cite{silaev2011microscopic}, where the material can have one coherence length larger and another smaller than the magnetic penetration depth. Over that, recently, a noticeable anisotropy in the upper critical field is reported\cite{bolotina2021checkerboard} in ZrB$_{12}$, which clearly depends on the orientation of the magnetic field vector. The authors associated the observation with two-gap superconductivity. However, such observation of anisotropy encourages us to testify the IT model proposed by Saraiva \textit{et al.}\cite{saraiva2019anisotropic} in ZrB$_{12}$. Also, as mentioned, the true order parameter symmetry of the superconducting gap for ZrB$_{12}$ is still debatable. Hence, an energy-resolved spectroscopic technique like PCARS will be further useful in this regard.


\section{Experimental Methods}


Both the surface-sensitive point contact Andreev reflection spectroscopy (PCARS)\cite{andreev1964thermal,naidyuk2005point} and bulk-sensitive ac susceptibility ($\chi$) measurements were performed using two different home-built probes inside the same liquid helium cryostat. For temperature-dependent experiments, a variable temperature insert (VTI) was used, the base temperature of which varies from 1.6 K to 2 K depending on different probes inside, and the helium level in the cryostat. For magnetic field dependent studies, a superconducting vector magnet is used, which can produce a maximum 6 T along the vertical direction parallel to the axis of the cryostat and 1 T each in two mutually perpendicular directions in the horizontal plane. The $\chi$ measurements were performed by sandwiching the sample between two coaxial copper coils connected to a lock-in amplifier. One coil was fed with the ac signal at 17.33 kHz frequency, and the other one was used as the pick-up coil. Magnetization and susceptibility (both ac and dc) measurements on ZrB$_{12}$ were reported multiple times in the past\cite{tsindlekht2004tunneling,daghero2004andreev,wang2005specific,leviev2005low,lortz2005specific,khasanov2005anomalous,tsindlekht2006glasslike,ge2014direct,ge2017paramagnetic} to characterize the material. Here, our objective of $\chi$ measurements was to verify the crystal quality and identify the transitions. As the exact value of $\chi$ is not of interest, $\chi$ is not volume corrected, and an arbitrary unit is used throughout. For the same reason, the imaginary and real components of $\chi$ are also not separated. For temperature and magnetic field dependence, the overall magnitude of $\chi$ is presented after a subtraction of the normal state saturation value. The point-contact probe works based on a differential screw mechanism and uses the standard needle-anvil method for contact formation. The surface of the crystal was properly cleaned before the spectroscopy experiments and the point contacts were formed \textit{in situ} at low temperatures with a Silver (Ag) tip. To have proper statistics, we probed different crystallographic facets of a single crystal of ZrB$_{12}$, and recorded spectra on different points on the same facet. The sample we have used for our studies was cut from a rod-shaped single crystal grown by floating-zone technique. One facet of the sample was cut along the crystal growth axis (c-axis). The spectra probed on that facet are described as `$z$-axis PCS' throughout our manuscript. The other facet was a plane perpendicular to the previous one. The spectra probed on this surface are described as `$x$-axis PCS'. For magnetic field dependent study of the spectra, the field was always applied perpendicular to the surface being probed. The same convention was used to describe the direction of the applied magnetic field in the susceptibility measurements.\\


\section{Directional Point Contact Andreev Reflection Spectroscopy}


Andreev reflection\cite{andreev1964thermal} is a quantum process that dominates the electronic transport through a ballistic point-contact between a normal metal and a superconductor. The process involves the reflection of a spin-up (down) electron as a spin-down (up) hole from the interface. This leads to a typical nonlinearity in the $I-V$ spectrum, which can be directly probed in a $dI/dV$ vs $V$ spectrum recorded across the point contact\cite{naidyuk2005point}. We employed a lock-in based modulation technique to probe such spectra and the modified Blonder-Tinkham-Klapwijk (BTK)\cite{plecenik1994finite} model to analyse the same. It may be noted that all PCAR data presented in this paper clearly show the hall-mark double-peak (symmetric about $V$ = 0) signature of Andreev reflection\cite{andreev1964thermal} and no other features like anomalous conductance dips\cite{sheet2004role}. This confirms that our measurements are performed in the spectroscopic (ballistic or diffusive) regime of transport. Compared to other superconductors we have studied earlier, it was surprisingly easy to find a contact in such spectroscopic regime for ZrB$_{12}$. The possible reasons can be the large coherence length(s)\cite{biswas2020coexistence,bolotina2021checkerboard}, an unusually high $T_c$ compared to other dodecaborides, and the high purity of the single crystal we have used of ZrB$_{12}$. As mentioned earlier, the extended flat DOS neighbourhood of the Fermi level\cite{shein2003band} for this material can be another possible reason behind such stability. The traditional BTK model\cite{blonder1982transition} assumes a $\delta$-function potential barrier whose strength is characterized by a dimensionless parameter $Z$. Apart from $Z$, the inelastic broadening parameter $\Gamma$ is used to take care of the finite quasi-particle lifetime in modified BTK theory\cite{plecenik1994finite}. The parameters $Z$, $\Gamma$ along with the superconducting energy gap $\Delta$ are used to fit a spectrum at a particular temperature $T$.\\


Fig. 1 represents PCAR spectra probed at $T$ $\sim$ 2 K on two distinct open facets of ZrB$_{12}$ and their statistics. Two typical spectra probed under the `$x$-axis PCS' configuration are shown with red circles in Fig. 1(a) and (b). The corresponding theoretical fits to the spectra using the modified BTK theory\cite{plecenik1994finite} are represented with black lines. The extracted fitting parameters $\Delta$, $\Gamma$ and $Z$ are also mentioned for each spectrum. A statistics of $\Delta$ for sixteen such independent spectra recorded on different points on the same surface, is presented in Fig. 1(c). Similarly, in Fig. 1(d) and (e), two typical spectra under `$z$-axis PCS' configuration are presented along with their corresponding theoretical fits and extracted parameters. A statistics of $\Delta$ for sixteen such independent spectra, all recorded on the same surface, is presented in Fig. 1(f). Some crucial observations are as follows.\\

For all spectra that we have analyzed, the normal state contact resistance varied between $\sim$ 0.9 to 4.5 $\Omega$. Using Wexler's formula\cite{wexler1966size} with these values, we found that the estimated contact diameter varied within a range from $\sim$ 12 to 27 nm. We did not find a direct correlation between the contact resistance $R_c$ and the measured order parameter $\Delta$. Furthermore, the spectra probed along the $z$-axis are visually more narrow and deep than those probed along the $x$-axis. Such spectra correspond to smaller values of the gap ($\Delta$). The median values of $\Delta$ are 0.87 meV and 0.72 meV for $x$-axis and $z$-axis contacts, respectively. Recently, two unique checkerboard charge stripe patterns were reported in ZrB$_{12}$\cite{bolotina2021checkerboard}. Static Jahn-Teller distortions in the crystal structure and a consequent anisotropic charge transfer that depends on different bond directions were argued as a possible mechanism behind that. Interestingly, in the same paper\cite{bolotina2021checkerboard}, the authors compared two different directions [100] and [110] and reported a signature of two-gap superconductivity (the ratios 2$\Delta_1/k_{B}T_{C}$ = 6 and 2$\Delta_2/k_{B}T_{C}$ = 2.5 respectively) based on their heat capacity study. First of all, a pair of gaps, if they are such widely separated in the energy scale, is expected to be resolved spectroscopically, particularly when the measurements are performed down to a temperature, that is $\sim$ 33$\%$ of $T_c$. In our data, no typical multigap features\cite{daghero2010probing} were visible. In our case, all spectra, irrespective of the surface it is probed on, are described well within the single gap $s$-wave BTK formalism\cite{blonder1982transition,plecenik1994finite}. Therefore, our measurements do not confirm the possibility of multigap superconductivity in ZrB$_{12}$. Nevertheless, as usual, the possibility of multiple gaps existing very close to each other in the energy cannot be strictly ruled out within the limits of the PCAR experiments.\\


In Fig. 2(a), we show the $T$ dependence of $\Delta$ as extracted from the spectral data presented in the inset. These data are for $x$-axis PCS. Similar $T$ dependence of the spectra for a $z$-axis PCS and the corresponding $T$ dependence of the extracted $\Delta$ are presented in Fig. 2(b). Modified BTK fit (black line)\cite{plecenik1994finite} for each spectrum is presented over the experimental data (colour circles). The expected temperature dependence for a conventional BCS superconductor\cite{bardeen1957theory} is presented as a dashed line along with each experimental $\Delta$ vs $T$ plot. The extrapolated zero temperature gap amplitude $\Delta(0)$ and the junction critical temperature $T_{c}^{j}$ are 0.85 meV (0.72 meV) and 5.8 K (5.9 K) for $x$-axis PCS ($z$-axis PCS), respectively. To note, although the superconducting gaps measured along the two different axes have differences in their spectral features and the extracted $\Delta$ values, they close nearly at the same temperature ($T_{c}^{j}$). Most importantly, in both cases, throughout the temperature range, the variations of $\Delta$s match extremely well with the corresponding BCS fits\cite{bardeen1957theory}. This further supports the conventional s-wave BCS superconductivity in ZrB$_{12}$ in contrast to some previous claims of d-wave symmetry\cite{gasparov2005electron,gasparov2006two} and pseudo-gap\cite{thakur2013complex} above $T_c$. Our observation of conventional superconductivity in ZrB$_{12}$ is also consistent with the previous PCARS measurements reported\cite{daghero2004andreev,girovsky2010strong} in this material. As we already have mentioned, the superconducting coupling strength of ZrB$_{12}$ varies widely among previous reports\cite{daghero2004andreev,tsindlekht2004tunneling,gasparov2004electron,khasanov2005anomalous,lortz2005specific,teyssier2007optical,gasparov2008study,girovsky2010strong}. From our spectral analysis, we found 2$\Delta/k_{B}T_{C}$ $\sim$ 3.4 for $x$-axis PCS and 2.8 for $z$-axis PCS, which indicate a weak coupling behaviour. The larger value matches with the previous reports by Lortz \textit{et al.}\cite{lortz2005specific}, Wang \textit{et al.}\cite{wang2005specific} and Teyssier \textit{et al}\cite{teyssier2007optical}. The smaller value agrees with the recent report by Bolotina \textit{et al.}\cite{bolotina2021checkerboard} corresponding to the smaller band in the two-band picture proposed by the authors.\\


So far, we have seen that the visual features and the corresponding gap values for spectra belonging to $x$-axis and $z$-axis PCS are different, but their $T_{c}^{j}$s are almost identical. This is not a surprise because $T_{c}$ is a global parameter\cite{bardeen1957theory} and, a unique superconductor ideally should have a unique $T_{c}$ unaffected by any anisotropy. However, as already discussed, the same restriction does not apply to the critical field, and to investigate further, we let the spectra probed on both surfaces evolve with an external magnetic field ($H$). In the upper inset of Fig. 3(a) we show the $H$ dependence of the $x$-axis PCS spectrum (also in Fig. 2(a)). With increasing $H$, the spectral features like the Andreev peaks and the spectral gap gradually become smaller, and disappear at $\sim$ 3 kG. A modified BTK fit\cite{plecenik1994finite} is also presented as a black line over the experimental data. The $H$ dependence of $\Delta$ extracted from the analysis is depicted in Fig. 3(a) which shows a smooth field dependence. However, the interface barrier strength $Z$, as extracted from such analysis\cite{plecenik1994finite} shows an unusual field dependence (lower inset in Fig. 3(a)). Specially for $x$-axis PCS, $Z$ appears to diverge as the field increases and approaches $H_{c}^{j}$. In the past such field dependent divergence of $Z$ was seen even in point contact with elemental Nb. Miyoshi $\textit{et al.}$ reported\cite{miyoshi2005andreev} such observation in Nb-Cu point contacts. It was shown that the divergence of $Z$ could be understood if the fitting model was modified to explicitly include the contribution of vortex cores in the experimentally measured conductance at finite magnetic field. Formation of vortex core at higher magnetic fields is indeed a likely scenario if the superconducting state is type-II in nature. We followed a similar treatment\cite{miyoshi2005andreev} and separated the contribution of the superconducting channel $(dI/dV)_{sc}$ from the experimental spectra ($dI/dV$) using the following relationship:\\

$dI/dV = h (dI/dV)_{vor} + (1-h) (dI/dV)_{sc}$\\

Here, $(dI/dV)_{vor}$ is the contribution of the normal channel due to the formation of vortex cores and $h=H/H_{c}^{j}$ with $H_{c}^{j}$ is 3 kG for $x$-axis PCS. The results of the fittings\cite{plecenik1994finite} on such extracted $(dI/dV)_{sc}$ are presented in Fig. 3(b). Beyond $\sim$ 2 kG the relative contribution of the superconducting channel becomes too small to have any meaningful fitting and extract reliable parameter values. Nevertheless, within the vortex core model, now $Z$ remains almost constant with increasing field strength for $x$-axis PCS, thereby removing the rather unphysical situation of diverging $Z$.\\

Now, let us concentrate on the $H$ dependence of the spectra for $z$-axis PCS. The initial spectrum (at $H$ = 0) considered here is the the same one that was used for the investigation of $T$ dependence in Fig. 2(b). The experimental spectra and the corresponding BTK fits\cite{plecenik1994finite} are presented in the upper inset of Fig. 4(a). The corresponding $H$ dependence of the extracted $\Delta$ is presented in Fig. 4(a) and $H$ dependent $\Gamma$ and $Z$ are presented in the lower inset. Interestingly, we found the evidence of anisotropy from the $H$ dependence of the spectra between $x$ and $z$ axis PCS. Though the field dependences of $\Delta$s visually look similar for both axes, the superconducting features appear to survive up to a higher field $\sim$ 6 kG for $z$-axis. Although for $z$-axis PCS no enhancement of $Z$ with magnetic field can be seen, for completeness, we also applied the `vortex core model'\cite{miyoshi2005andreev} with $H_{c}^{j}$ = 6 kG for $z$-axis PCS. The results of such analysis are presented in Fig. 4(b). Beyond $\sim$ 2 kG, no meaningful fittings seem to be possible due to high noise in the extracted $(dI/dV)_{sc}$. No noticeable change in the results was seen after applying the vortex core based model for the $z$-axis PCS. From this analysis, we draw two important conclusions. First, we could understand the enhancement of $Z$ with $H$ for the $x$-axis spectra within a vortex core model\cite{miyoshi2005andreev}. Such enhancement was not seen for the $z$-axis spectra and hence a `vortex core model' is not relevant in that case. When we still applied the model by brute force for $z$-axis spectra, as expected, no significant change in the $H$ dependence of $Z$ was seen. This observation hints to the possibility of the formation of vortex cores when the field is applied along $x$-axis, while the existence of vortices is not guaranteed (within the limit of our results) for the field applied along $z$-axis. Second, the spectral gap $\Delta$ sharply falls within a field range of 0.5 to 1.5 kG for both x (Fig. 3(b)) and z axis (Fig. 4(b)) PCS. This field scale is similar to the reported values of upper critical field $H_{c2}(0)$ (550 G - 1.5 kG)\cite{wang2005specific,daghero2004andreev,gasparov2005electron} for bulk ZrB$_{12}$. However, we also note that the local critical field $H_{c(l)}$ measured from the PCARS experiments, defined as the field where the gap feature completely vanishes, is much larger than 1.5 kG irrespective of the direction of the applied magnetic field. In fact, the critical fields thus measured are an order of magnitude larger than the the bulk critical field (as reported by other groups\cite{tsindlekht2004tunneling,wang2005specific,lortz2005specific} and also from our $\chi$ measurements to be discussed later) of ZrB$_{12}$. According to previous reports, $H_{c2}(0)$ varies widely based on the different measurement techniques employed. These aspects of critical field variation were attributed to a two-gap scenario by
Bolotina \textit{et al.}\cite{bolotina2021checkerboard}. To resolve the puzzle regarding the electron-phonon coupling strength in this material, some authors also have proposed a dramatically enhanced $H_{c3}(0)$ within a framework of surface enhancement of superconductivity\cite{wang2005specific,tsindlekht2004tunneling}. Under such circumstances, PCAR experiments are expected to be influenced by the enhanced surface critical field. Also the factors like the confined dimension of the point contact geometry and the enhanced local disorder under a point contact can contribute to the difference between $H_{c(l)}$ and $H_c$. In experiments, either all or some of these factors lead to the measurement of an enhanced critical field on superconducting point contacts. In order to further investigate the critical field anisotropy in this system, we investigated the anisotropy of the screening properties with magnetic field angle as discussed below.

\section{AC Magnetic Susceptibility through Two-coil Mutual Inductance Measurements}


Temperature-dependent $\chi$ measured by two-coils setup shows a sharp superconducting transition at $T$ = 5.9 K. This transition temperature ($T_c$) is consistent with the previous reports\cite{matthias1968superconductivity,gasparov2004electron,daghero2004andreev,lortz2005specific,wang2005specific,khasanov2005anomalous,gasparov2005electron,gasparov2006two}, and such a sharp transition supports the high purity of the single crystal we are using. In Fig. 5(a), the magnetic field dependence of such transition is shown. The static magnetic field is applied along the $z$-axis and increased with 50 Gauss steps. With the increasing magnetic field, a clear shift of the $T_c$ towards lower temperatures is visible. Each transition is recorded during zero-field cooled heating condition. Interestingly, at higher field, a kink structure appears at the onset part of the transition, which seemingly breaks the transition curves in two parts. The possible reason behind such kink is two independent transitions in the real (in phase) and imaginary (out of phase) parts of $\chi$\cite{daghero2004andreev}.\\


In Fig. 5(b), we show the magnetic field dependence of the $\chi$ using the same method as before with temperature kept fixed at $T$ = 1.6 K. When the magnetic field is applied along the $z$-axis, the superconducting transition starts at $H$ $\sim$ 450 G and completes at $H$ $\sim$ 560 G. No hysteresis can be observed in this case. An entirely different result is observed when the magnetic field is applied along the x or y direction. For a clearer view in Fig. 5(b), plots for the $H$ $\parallel$ y-axis and $H$ $\parallel$ $z$-axis are vertically shifted, keeping the same scaling. Transition starts at $H$ $\sim$ 485 $\pm$ 25 G and completes at $H$ $\sim$ 645 $\pm$ 25 G for both x and y directions. The higher values are for increasing field, and the lower values are for decreasing field, respectively. A clear hysteresis is observed in both cases. As we have already mentioned, the reported thermodynamic critical field $H_c(0)$ of ZrB$_{12}$ varies widely within a range of 300 - 415 G in literature\cite{tsindlekht2004tunneling,wang2005specific,lortz2005specific}. Moreover, the determination of exact $H_c(0)$ based on an ac technique is difficult because the transition curve highly depends on both the amplitude and frequency of the signal\cite{tsindlekht2004tunneling,tsindlekht2006glasslike}. Nevertheless, that dependence should not affect our observation of anisotropy because we kept the ac signal constant in the primary coil and changed only the temperature and the externally applied field as tuning parameters.\\

Hysteretic behaviour in ZrB$_{12}$ was seen in different contexts in the past too. For example, Tsindlekht \textit{et al.}\cite{tsindlekht2004tunneling} mentioned about a tiny hysteresis loop they found near the critical field in their dc magnetization measurements but did not present that in their data. Based on their observation, the authors also concluded that the bulk ZrB$_{12}$ could be either a type-I or a type-II superconductor as it has a marginal value of $\kappa$ $\sim$ 0.71. Wang \textit{et al.}\cite{wang2005specific} also reported hysteresis both in field dependent specific heat and magnetization measurements within type-II/I temperature region. To note, intermediate state in Type-I superconductor can also trap magnetic field in the similar way. But, the intermediate regions are macroscopic and will trap field applied from any direction. The observed hysteresis for $H$ $\parallel$ x and $H$ $\parallel$ y-axis can be attributed to formation of vortices and trapping-de-trapping phenomena of the same. The absence of hysteresis for $H$ $\parallel$ $z$-axis, on the other hand, may indicate that such vortex physics is absent for the given field direction. Therefore, we found indication of the possibility of a mixed state of type-II superconductivity for $H$ $\parallel$ x and $H$ $\parallel$ y-axis, and type-I superconductivity for $H$ $\parallel$ $z$-axis in this material. Here we also note that, from our PCARS measurements, we found an anomalous, diverging magnetic field dependence of parameter $Z$ for $x$-axis point contact which could be ultimately resolved when the contribution of the vortex channel to the conductance was excluded before analysis. Such diverging $Z$ with magnetic field was not seen for $z$-axis PCS. This is also consistent with the idea of type-II superconductivity for $H$ $\parallel$ x and y-axis but type-I superconductivity for $H$ $\parallel$ $z$-axis in ZrB$_{12}$. Recently Saraiva \textit{et al.}\cite{saraiva2019anisotropic} have theoretically shown that the idea of inter-type superconductor (as we discussed in the introduction) can also be applied for anisotropic superconductors. To make this work, an appropriate scaling transformation\cite{klemm1980lower,blatter1992isotropic} have to be introduced in the formalism describing the anisotropic superconductor. Such a transformation depends on the direction of the applied magnetic field. Consequently, the type of superconductivity may be different for the magnetic field applied along different directions. This is precisely what we observed experimentally in ZrB$_{12}$.\\

In case of type-I superconductors, it is often expected that the superconducting gap would show a sharp first-order disappearance with increasing magnetic field. Such an effect was earlier seen in the type-I phase of a number of superconductors\cite{naidyuk1996temperature,naidyuk1996magnetic,le2019single}. On the other hand, the observation of the first-order destruction of superconductivity is often limited in PCAR experiments because the presence of the non-superconducting tip may cause an enhanced local disorder, specially for the fact that a point contact is effectively a mesoscopic entity with confinement. Enhancement of local disorder decreases the local mean free path, which in turn reduces the local coherence length of the superconducting fraction of the point contact, thereby driving the superconductor locally to the type-II regime (large G-L parameter $\kappa$). Consequently, even a strictly type-I superconductor like Pb often loses its type-I behaviour in mesoscopic dimensions\cite{yang2003enhanced,he2013giant,rodrigo2012topological,sirohi2016transport}. For ZrB$_{12}$, as $\kappa$ is very close to the critical value 1/$\sqrt{2}$\cite{wang2005specific}, a small disorder introduced by the tip is enough to drive the system to the type-II regime locally. We believe that this is why we did not observe the sudden destruction of the superconducting gap in ZrB$_{12}$ with field previously for the $z$-axis point contacts.\\


To investigate the anisotropic behaviour further, we performed field angle dependence of $\chi$. We chose six unique points (P$_{1}$ - P$_{6}$) from the $\chi$ vs $H$ plot for $H$ $\parallel$ $z$-axis in Fig. 5(b). As described in Fig. 5(c), six such points correspond to six different magnetic field values (H$_{1}$ - H$_{6}$). For each point, we kept the magnitude of the magnetic field fixed but rotated the direction of the field in the x-z plane. The temperature was kept fixed throughout the experiment at 1.6 K. In Fig. 5(d), we show the angular dependence of $\chi$ corresponding to these six field values H$_{1}$ to H$_{6}$. Angle $\theta$ is measured \textit{w.r.t} $z$-axis in the z-x plane. Plots for higher fields are given vertical shifts for an uncluttered view. At H$_{1}$ = 200 G, the material is deep inside the superconducting domain and a flat angular dependence of $\chi$ is visible. Similar flat dependence is visible at H$_{6}$ = 920 G when the sample is in the normal state. From H$_{2}$ = 400 G to H$_{5}$ = 760 G, when the material passes through the superconducting transition, a clear two-fold symmetry appears and then disappears with increasing $H$. The maximum anisotropy was found at H$_{4}$ = 560 G. At this field $H$ $\parallel$ $z$-axis, the normal state is just critically reached (see point P$_{4}$ in Fig. 5(c)).  However, the material is still superconducting at the same field value when $H$ $\parallel$ $x$-axis (or y-axis), as shown in Fig. 5(b). The directional PCARS and field-angle dependent $\chi$ measurements collectively confirm that the observed anisotropic behaviours in ZrB$_{12}$ does not originate from extrinsic geometry-driven effects, but is indeed an intrinsic property of the material system.

\section{Conclusion}

We explored the anisotropy in the superconducting properties of ZrB$_{12}$ by point contact Andreev reflection spectroscopy and ac susceptibility experiments. Based on our PCARS experiments, we recorded 2$\Delta/k_{B}T_{c}$ $\sim$ 2.8 and $H_{cl}$ $\sim$ 6 kG along a primary axis (001) of the single crystal, while in the perpendicular direction we recorded 2$\Delta/k_{B}T_{c}$ $\sim$ 3.4 and $H_{cl}$ $\sim$ 3 kG. The spectra fit nicely with a single-gap s-wave BTK model\cite{blonder1982transition,plecenik1994finite} and corresponding $\Delta$s follow BCS temperature dependence\cite{bardeen1957theory} in both cases. Our detailed analysis on magnetic field dependent spectra indicated the presence of vortex cores along the (001) direction but not in the perpendicular direction. From our ac susceptibility measurements, we found a non-hysteretic (type-I like) transition for the field applied along the (001) direction but a hysteretic (type-II like) transition in perpendicular directions. An angular two-fold symmetry in the magnetic field dependence was also observed from the same experiment. Based on these observations, we surmise that the superconducting ZrB$_{12}$ has a magnetic anisotropy, which can be used to manipulate its superconducting state as well as its superconducting type by changing the direction of the applied field. This is the core argument behind the inter-type formalism developed by Saraiva \textit{et al.}\cite{saraiva2019anisotropic} for an anisotropic superconductor. Here, using ZrB$_{12}$ as a model system, we have provided the experimental evidence of the same. Our findings will also be insightful to explore the controlled magnetic behaviours of other superconductors with marginal Ginzburg-Landau parameter.

\section{Acknowledgements}

S.D thanks Ranjani Ramachandran for her help with the experiments, and Dr. Shirshendu Gayen for his contribution in the instrument interfacing and data fitting programmes. R.P.S acknowledges the financial support from the Science and Engineering Research Board (SERB)-Core Research Grant (grant No. CRG/2019/001028). G.S acknowledges financial support from the Swarnajayanti fellowship awarded by the Department of Science and Technology (DST), Govt. of India (grant No. DST/SJF/PSA-01/2015-16).





\bibliography{Bibliography}	

\begin{figure}[htbp]
	\centering
	\includegraphics[width=0.85\textwidth]{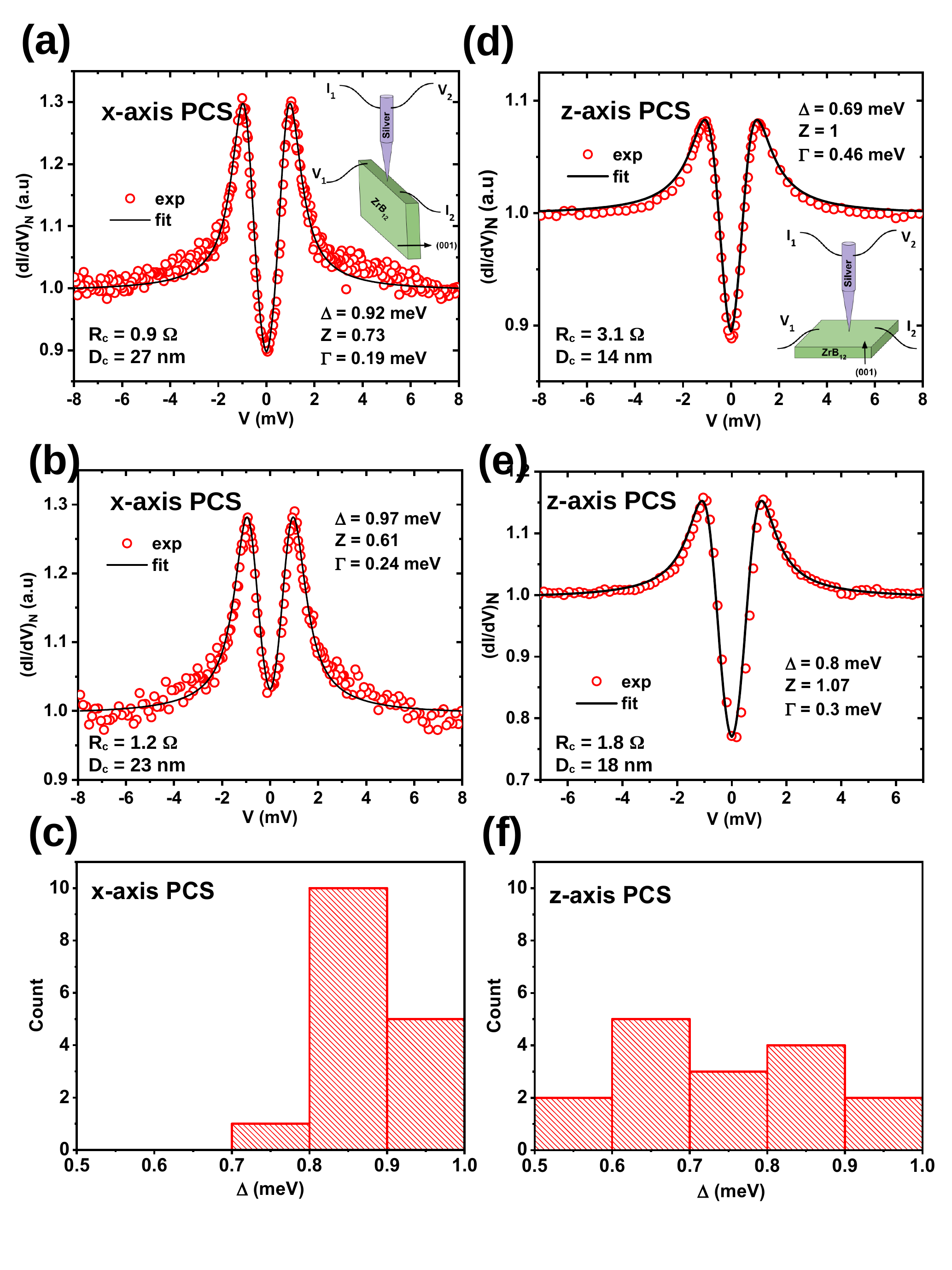}
	\caption{Representative conductance spectra (red circles) for point contacts formed along $x$-axis (\textbf{(a)}, \textbf{(b)}) and along $z$-axis (\textbf{(d)}, \textbf{(e)}). Statistics of the fitting parameter $\Delta$ for 16 independent point contact spectra formed along $x$-axis (\textbf{(c)}) and the same for $z$-axis (\textbf{(f)}). All spectra are recorded at temperature ($T$) $\sim$ 2 K and in absence of any applied magnetic field.}
	
\end{figure}

\begin{figure}[htbp]
	\centering
	\includegraphics[width=0.75\textwidth]{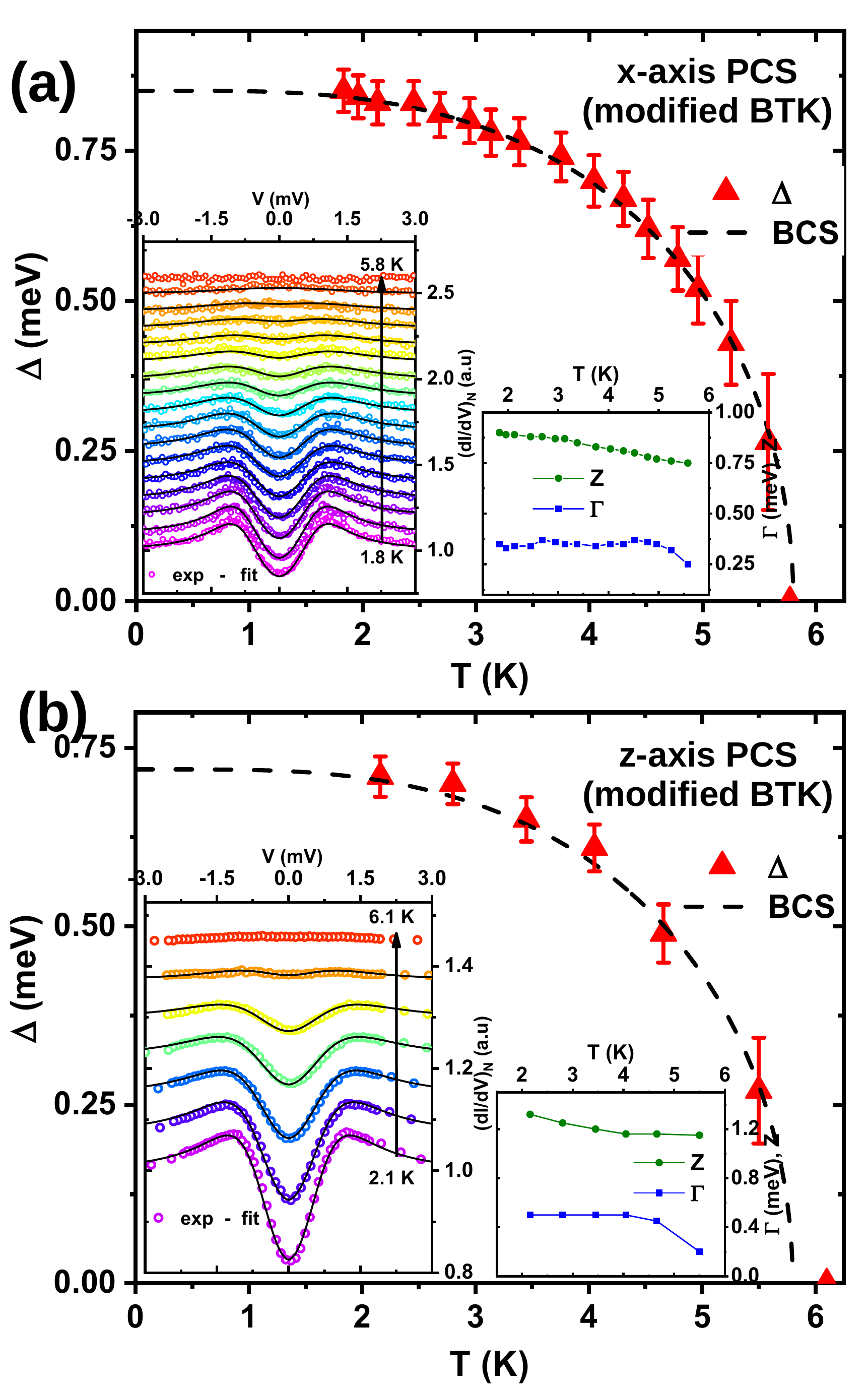}
	\caption{Evolution of the superconducting gap ($\Delta$) with temperature ($T$). Upper inset: The $T$ dependence of the conductance spectra and corresponding BTK fits. Lower inset: The $T$ dependences of parameters $\Gamma$ and $Z$. \textbf{(a)} For $x$-axis PCS and \textbf{(b)} for $z$-axis PCS.}
		
\end{figure}

\begin{figure}[htbp]
	\centering
	\includegraphics[width=0.75\textwidth]{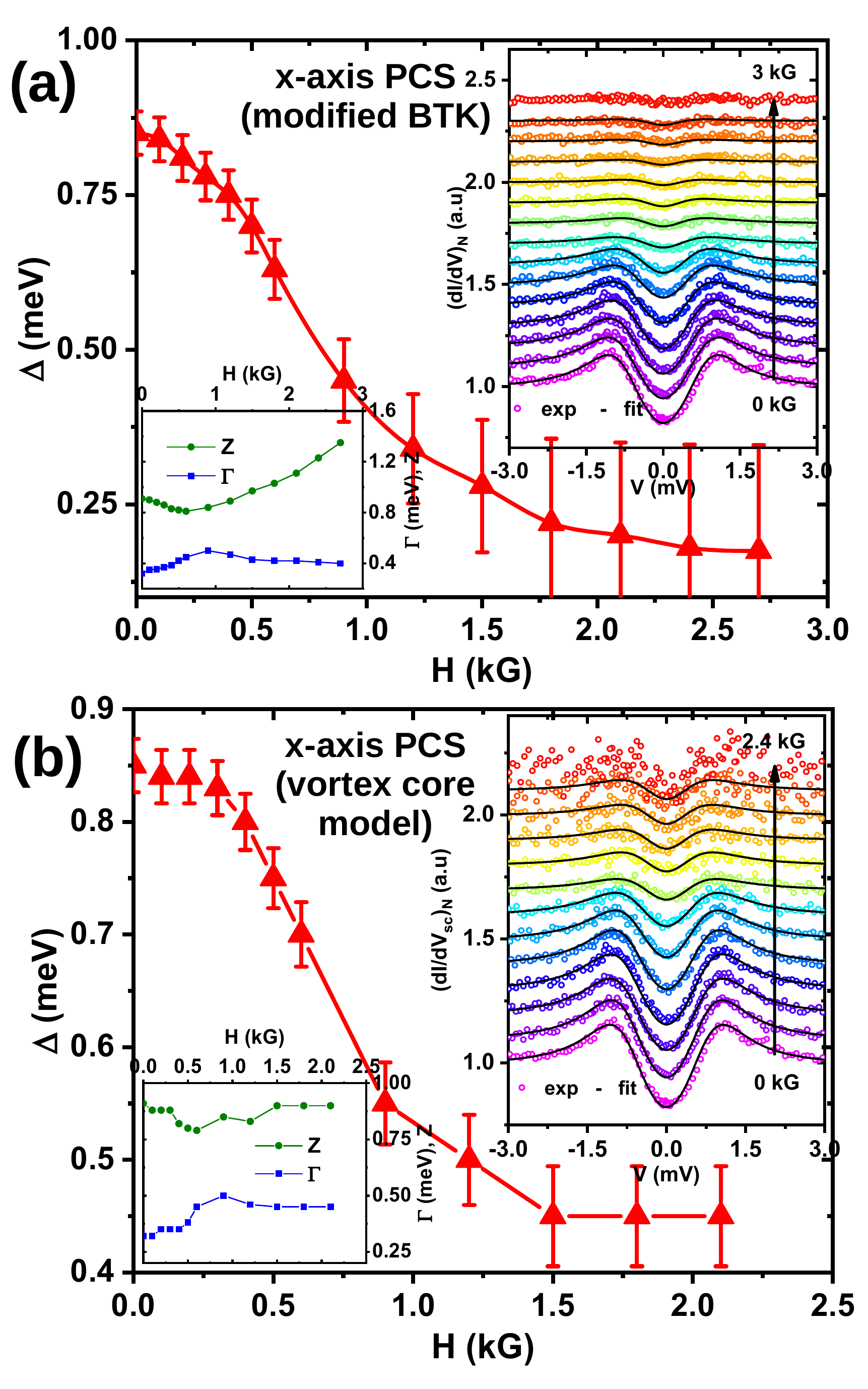}
	\caption{\textbf{(a)} Evolution of $\Delta$ with external magnetic field ($H$) for $x$-axis PCS. Upper inset: The $H$ dependence of the experimental $dI/dV$ with modified BTK fits. Lower inset: The $H$ dependences of parameters $\Gamma$ and Z. \textbf{(b)} Similar field dependences of the parameters after subtracting the contribution of vortex core from the experimental $dI/dV$.}
	
\end{figure}

\begin{figure}[htbp]
	\centering
	\includegraphics[width=0.75\textwidth]{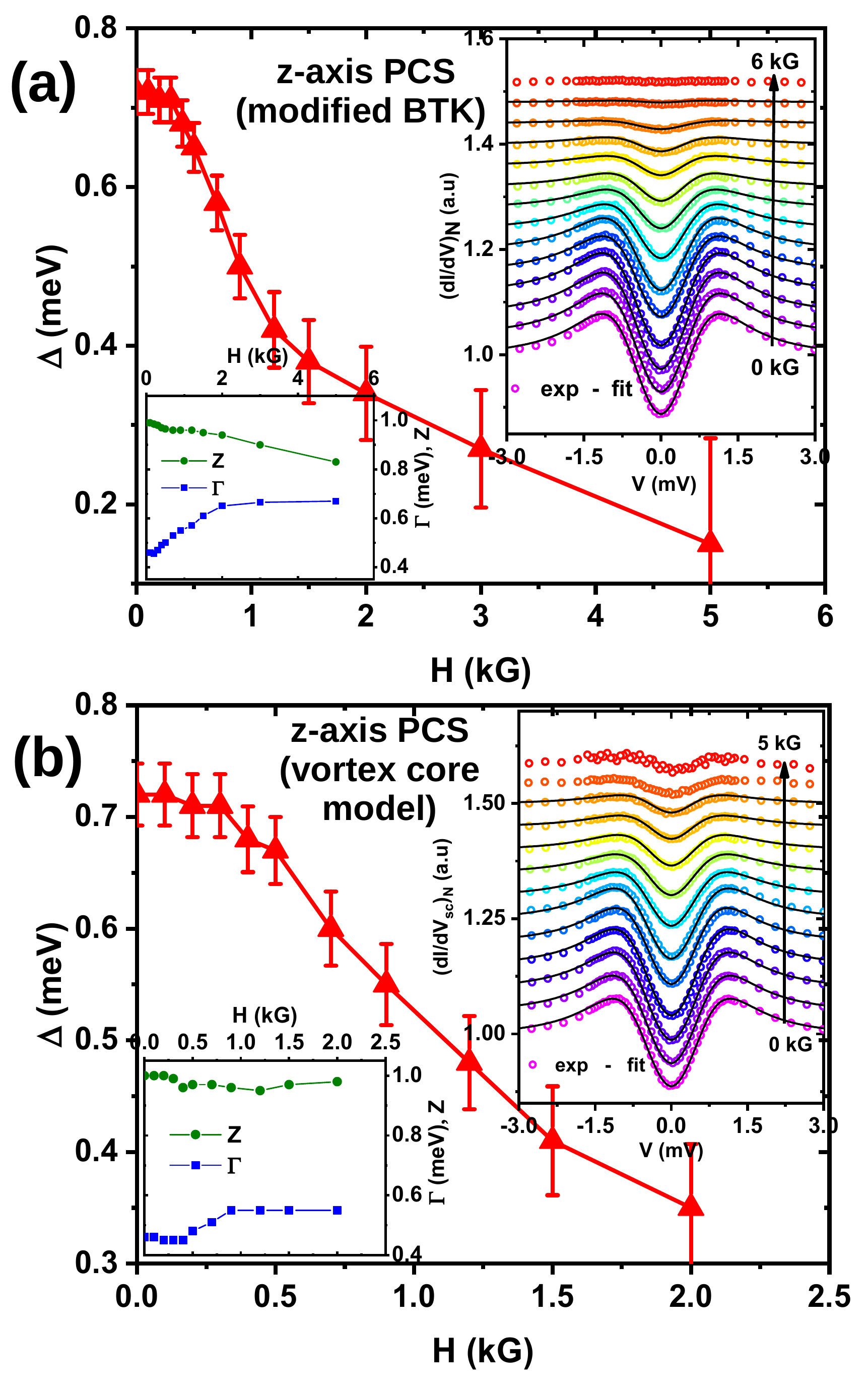}
	\caption{\textbf{(a)} Evolution of $\Delta$ with $H$ for $z$-axis PCS. Upper inset: The $H$ dependence of the experimental $dI/dV$ with modified BTK fits. Lower inset: The $H$ dependences of parameters $\Gamma$ and $Z$. \textbf{(b)} Similar field dependences of the parameters after subtracting the contribution of vortex core from the experimental $dI/dV$.}
	
\end{figure}

\begin{figure*}[htbp]
	\centering
	\includegraphics[width=1\textwidth]{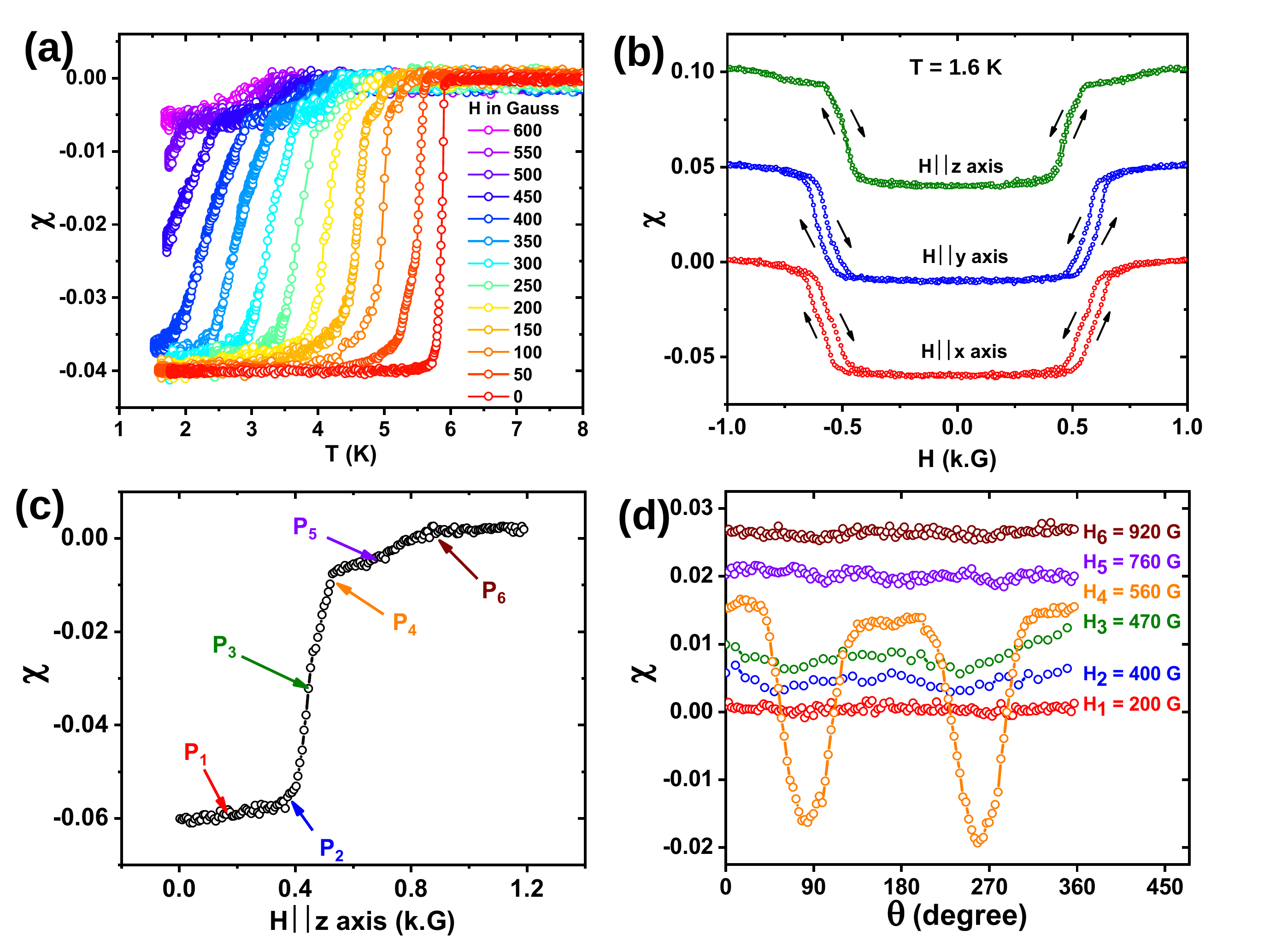}
	\caption{\textbf{(a)} External magnetic field ($H$) dependence of the bulk $T_c$ from two-coil mutual inductance measurements. All data are taken during zero-field cooled warming and the ac magnetic susceptibility ($\chi$) is not volume corrected. \textbf{(b)} Anisotropy in the $H$ dependence of $\chi$ showing hysteresis when $H$ $\parallel$ $x$-axis or $H$ $\parallel$ y-axis but no hysteresis when $H$ $\parallel$ $z$-axis. \textbf{(c)} Six points (P$_{1}$ - P$_{6}$) are chosen on the right half (+ive $H$) of $H$ $\parallel$ $z$-axis plot in (b). \textbf{(d)} Angular dependence of $\chi$ with six different values of magnetic fields (H$_{1}$ - H$_{6}$) corresponding to the six points (P$_{1}$ - P$_{6}$) chosen in (c). Plots in (b) and (d) are given vertical shifts keeping same scaling for a clearer view.}
	
\end{figure*}

\end{document}